\documentclass[%
     reprint,
     amsmath,amssymb,
     aps,
    ]{revtex4-1}
    
    \usepackage{graphicx}
    \usepackage{dcolumn}
    \usepackage{bm}
    \usepackage{hyperref}
    \usepackage{amsmath} 
    \usepackage[version=4]{mhchem} 
    
    \usepackage{color}
    
\begin{document}
    
    \preprint{APS/123-QED}
    
    \title[Neutrino SNSI]{Bounds on neutrino-scalar nonstandard interactions from big bang nucleosynthesis}
    
    \author{Jorge Venzor}\email{jorge.venzor@cinvestav.mx}
    \author{Abdel P\'erez-Lorenzana}\email{aplorenz@fis.cinvestav.mx}
    \affiliation{ Departamento de F\'{\i}sica, Centro de Investigaci\'on y de Estudios Avanzados del I.P.N. 
                    Apdo. Post. 14-740, 07000, Ciudad de M\'exico, M\'exico. }
    \author{Josue De-Santiago}\email{Josue.desantiago@cinvestav.mx}
    \affiliation{ Departamento de F\'{\i}sica, Centro de Investigaci\'on y de Estudios Avanzados del I.P.N. 
                    Apdo. Post. 14-740, 07000, Ciudad de M\'exico, M\'exico. }
    \affiliation{C\'atedra - Consejo Nacional de Ciencia y Tecnolog\'{\i}a.
    			Av. Insurgentes Sur 1582, 03940, Ciudad de M\'exico, M\'exico}

    
    \begin{abstract}
    
Coherent forward scattering processes by neutrino-scalar nonstandard interactions (SNSI) induce an effective neutrino mass. In the Early Universe, a large neutrino effective mass restricts the production of neutrinos. The SNSI effect is modulated by two effective couplings, these account for the coupling between neutrinos and electrons/positrons, $G_{\rm eff}$, and the neutrino self-interaction, $G_{\rm S}$. These parameters are directly related to the effective number of relativistic species and non-zero values imply a smaller than expected $N_{\rm eff}$. We employ big bang nucleosynthesis to constraint the SNSI effect. We find that $ G_{\rm eff }< 1.2$ MeV$^{-2}$ and $ G_{\rm S }< 2.0 \times 10^{7}$ MeV$^{-2}$ at 68\% CL. For a scalar mass in the range $10^{-15} {\rm eV}\lesssim m_{\phi}\lesssim 10^{-5}{\rm eV}$, our neutrino-scalar coupling constraint is more restrictive than any previous result.
    \end{abstract}
    
    \keywords{Cosmological neutrinos, Big bang nucleosynthesis, Cosmology, Cosmic microwave background, Solar neutrinos}
    \maketitle

    \section{\label{sec:introduction} Introduction}
    
    Despite being light, neutrino gravitational interaction plays an essential role in shaping the distribution of matter and energy in the Universe.
    Several cosmological surveys have led to the strongest bounds on the sum of the neutrino masses \cite{Aghanim2018Planck,Palanque_Delabrouille_2015,Loureiro2019,Gariazzo2018}.
    These are one order of magnitude better than those from experimental counterparts \cite{KATRIN2019}.
    Cosmology now leads the race to determine the neutrino mass hierarchy, and possibly, measure the mass of at least one neutrino throughout this decade \cite{Brinckmann2019,Archidiacono2020}.
    Moreover, three standard neutrinos are required to predict accurately the abundance of light elements on the Universe through big bang nucleosynthesis (BBN) \cite{Cooke2018,Aver2015,Peimbert2016}.
    This is in concordance with the standard precision computation of the neutrino contribution to radiation density, that can be expressed in terms of the parameter $N_{{\rm eff}}\simeq 3.046$ \cite{Mangano2005,deSalas2016,Escudero2020,Akita2020,froustey2020neff}.

    Cosmological model-independent bounds on neutrinos will be more reliable by disentangling the effects of neutrino parameters with the rest of cosmological ones \cite{Aghanim2018Planck}.
    As an important step, the existence of relativistic species in the Early Universe has been proven by detecting a phase shift on the acoustic oscillations that cannot be mimicked by other cosmological parameters \cite{Bashinsky2004,Follin2015,baumann2019first}.
    In this sense, cosmology has become a fruitful \textit{Lab} to test neutrino physics in the outline of the standard model of particle physics (SM) and beyond (BSM).

    Neutrino interactions with matter are crucial to study them. 
    For instance, the Mikheyev Smirnov Wolfenstein (MSW) effect \cite{Wolfstein1979}, which changes the neutrino oscillations in matter, was used to determine the sign of the square-mass splitting $\Delta m_{21}^2>0$ in the solar neutrino experiments (see for instance \cite{SNO2004,Borexino2011}).
    The same mechanism is being brought out by long-baseline neutrino experiments aiming to determine the sign of $\Delta m_{31}^2$ (see experiments \cite{NOvA2019,T2K2020}).
    
    In cosmology, a neutrino nonstandard interaction (NSI) may solve some tensions in the standard theory.
    It has been studied whether an NSI may explain the discrepancy known as the $H_0$-tension, where the measurement of $H_0$ by the cosmic microwave background (CMB) and local observations are clearly in statistical
    disagreement \cite{H0licow2020,Riess2019ApJ,Verde2019,Aghanim2018Planck}.
    There are some approaches that try to solve this problem using NSI,  including interactions in the sterile  \cite{Hannestad2014SI,*Dasgupta2014,*Archidiacono2015Hannestad,*Archidiacono2016JCAP,*Archidiacono2016PRD,*Chu2018,*archidiacono2020sterile},
    or in the active neutrino sectors  \cite{Bell2006,*Archidiacono2014,*CyrRacine2014,*Oldengott2015,*Lancaster2017,*Oldengott2017,*Kreisch2020}.
    In the latter approach, neutrinos are required to be either strongly self-interacting (SI$\nu$) or moderately self-interacting (MI$\nu$).

    SI$\nu$ and/or MI$\nu$ are assumed to be mediated by a scalar particle with a mass larger than $\mathcal{O}(\rm{keV})$.
    And, the parameter space, in this approximation, has been cornered by experimental, astrophysical, and BBN constraints \cite{Heurtier2017,farzan2018,Huang2018,Blinov2019,Brune2019}.
    On the other hand, the phenomenology of neutrino scalar nonstandard interactions (SNSI) mediated by a light particle is rich and has several consequences.
    For instance, large-scale structure (LSS) data constrains neutrino dispersion mediated by a scalar much lighter than $\mathcal{O}(\rm{eV})$  \cite{Forastieri2015,Forastieri2019}.
    Furthermore, neutrinos may annihilate and decay into lighter bosons, which, interestingly, may relax the bound on $\sum m_{\nu}$ imposed by LSS \cite{Beacom2004,Hannestad2005,Escudero2019,Chacko2020,Escudero2020relaxing}.
    
    Although the information on the light mediator mass is lost when studying two-body dispersion in the regime $m_{\phi}\ll T_{\nu}$, loop diagrams such as mass-correction type, a priory, are mass-dependent regardless of the smallness of the scalar mass. 
    Therefore, studying this kind of diagrams within the Early Universe background is convenient if we are to search for mediator mass-dependent SNSI constraints.
  
    In this manuscript, we explore the cosmological consequences of neutrino SNSI mass correction processes mediated by a light scalar particle ($10^{-15} {\rm eV}\lesssim m_{\phi}\lesssim 10^{-5}{\rm eV}$ ).
    We assess the calculations performed by Babu \textit{et al.} \cite{Babu2020scalar} in the Early Universe.
    Mass correction diagrams involving an SNSI have received recent attention because Ge \& Park \cite{Ge2019} found a small solar neutrino data preference for non-vanishing SNSI couplings
    with ordinary matter.
    This result has led to further research about neutrino propagation with SNSI in The Earth, The Sun, and supernovae \cite{Smirnov2019,Babu2020scalar,Khan2020}.

    For our exploration, we identify two effective parameters that modulate the SNSI effect and study its consequences.
    We solve numerically the mass contribution and the evolution of the neutrino density. 
    Additionally, we notice that large effective SNSI couplings may noticeably change the neutrino contribution to radiation.
    This information is encoded through a temperature-dependent change on the effective number of relativistic species $N_{{\rm eff}}$.
    In order to find $N_{{\rm eff}}$, we employ a modified version of the public code \textsc{nudec\_bsm} \cite{Escudero2020,Escudero2019_Neutrino_decoupling_BSM}.
    A change on $N_{{\rm eff}}$ straightforwardly alters the expansion rate during radiation dominated era, affecting the proton$\leftrightarrow$neutron freeze-out temperature and, hence, the neutron to proton ratio right at the unset of BBN.
    Thus, the production of primordial nuclei helps us to constrain the SNSI parameter space.
    We use a modified version of the public BBN code \textsc{alterbbn} \cite{Arbey2012,AlterBBNv2} to find the parameter constraints.
    Finally, we translate these bounds into the scalar mass - couplings parameter space and compare them with other results.

    The rest of the paper is organized as follows. In Section \ref{sec:nsi} we review and discuss the properties of the effective mass coherent forward scattering process (CFS) by SNSI at high temperatures.
    In Section \ref{sec:cosmo_imp} we explore the phenomenological consequences of the effective neutrino mass.
    Then, in Section \ref{sec:BBN} we constrain the parameter space of the SNSI with BBN theory and the abundance of light elements.
    In Section \ref{sec:comparison} we compare our constraints with laboratory, astrophysical and cosmological bounds on the parameters.
    Our conclusions are summarized in Section \ref{sec:conclusions}.
    
    
    \section{\label{sec:nsi} Neutrino scalar nonstandard interactions}

    \begin{figure}
    \centering
    \includegraphics[width=0.4\textwidth]{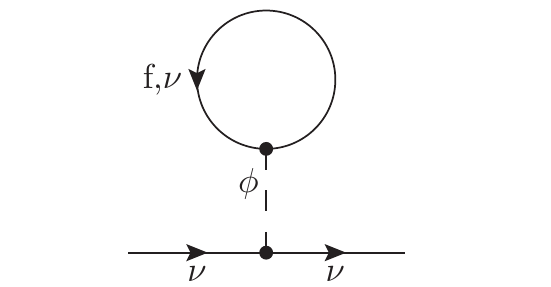}
    \caption{\label{fig:tadpole} Tadpole diagram of the neutrino CFS with a background of leptons. This process induces a thermal correction to neutrino mass.}
    \end{figure}

    The outcomes of neutrino NSI depend on the nature of the mediator particle.
    On the one hand side, vector-mediated NSI has a phenomenology that produces similar effects as the weak interaction.
    The SNSI instead, appears as a Yukawa term on the effective Lagrangian \cite{Ge2019} which induces an effective mass.
    This mass term depends on the properties of the environment where neutrinos propagate.
    A dense and hot background may produce a large neutrino mass.

    We are interested in the effect of the SNSI in the CFS described by the tadpole diagram of Fig. \ref{fig:tadpole}.
    We consider that neutrinos are propagating in a hot plasma when the Universe had a temperature around some MeVs, this plasma is composed of photons, baryons, charged leptons, and the three standard neutrinos.
    The SNSI effect in the neutrino propagation can be interpreted as a refractive index \cite{Liu1992,peltoniemi1999coherence}.
    Here, we focus on a generic scalar interaction ignoring the details of an underlying particle physics model, having the cosmological phenomenology as our main approach.
    
    The effective neutrino mass described by the quantum correction would be
    \begin{equation}\label{eq:meff}
        m_{{\rm eff}}=m_{\nu} + 2 G_{{\rm eff}}\Delta m (m_e;T_{\gamma}) + 3  G_{{\rm S}}\Delta m (m_{\nu};T_{\nu}),
    \end{equation}
    where $m_{\nu}$ is the bare neutrino mass, and the correction is described by \cite{Babu2020scalar}
    \begin{equation}\label{eq:deltam}
        \Delta m (m_f;T) = \frac{m_f}{\pi^2}\int_{m_f}^{\infty}dk \sqrt{k^2-m_f^2} f(k).
    \end{equation}
    Here $m_f$ is the mass of the fermion and $f(k_0)$ is the Fermi-Dirac distribution for the the background fermions.
    Safely neglecting the chemical potential \cite{Thomas2020}, $\mu=0$, the Fermi-Dirac distribution for both fermions and anti-fermions is the same $({\rm e}^{k/T}+1)^{-1}$, where $T$ is the temperature of the thermal background.
    The two free parameters, $G_{{\rm eff}}$ and $G_{{\rm S}}$, are then given as
    \begin{equation}
    G_{{\rm eff}} = \frac{g_f g_{\nu}}{m_{\phi}^2},
    \end{equation}
    and
    \begin{equation}
    G_{{\rm S}} = \frac{g_{\nu}^2}{m_{\phi}^2},
    \end{equation}
    where $m_{\phi}$ is the the mass of the scalar mediator, $g_{\nu}$ is the neutrino-scalar coupling and $g_f$ is the coupling between the scalar and charged leptons.
    These effective couplings encode the strength of the interaction and are the ones to be constrained by observations.
    Here we assume universal couplings with both charged lepton and neutrino flavors.
    Therefore, all complex phases can be absorbed and one can assume neutrino mass corrections to be always positive. 
    Notice that, at the temperatures that we are interested here, there are not muons/taus present in the plasma, since they have already decayed into lighter particles by then ($T_{\gamma}\ll m_{\mu}\sim 105.65 $ MeV).
    Hence, we only take into account couplings with electrons and positrons.

\begin{figure}
    \centering
    \includegraphics[width=0.48\textwidth]{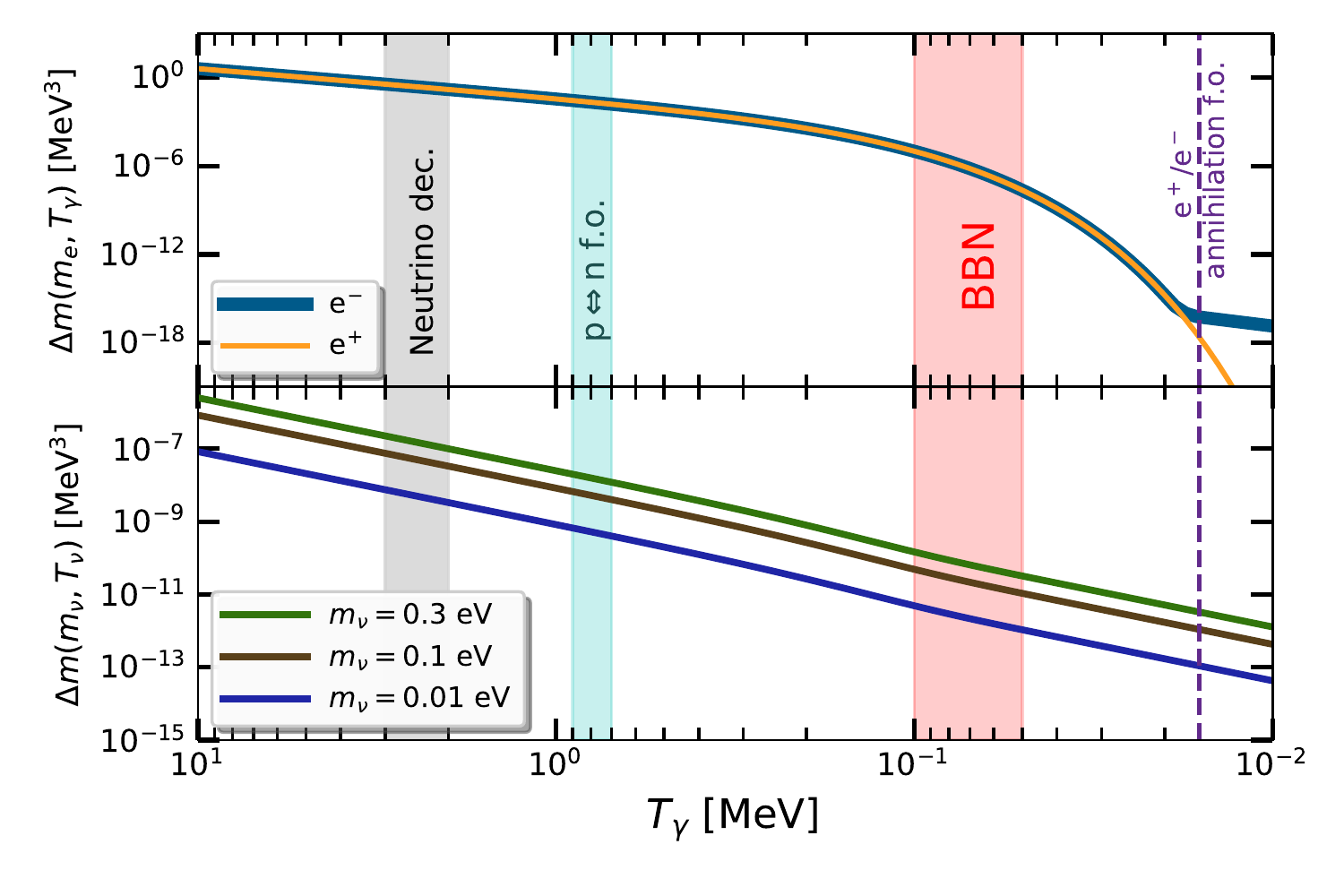}
    \caption{\label{fig:mass_correction}
    Neutrino mass-correction induced by an SNSI interaction, depicted as a function of the photon-baryon temperature. \textit{Upper panel}: SNSI with electrons and positrons. \textit{Lower panel}: Neutrino self-interaction for three different values of the neutrino bare mass present in the background.  Key cosmological events highlighted: Neutrino decoupling, proton to neutron freeze-out (f.o.), synthesis of light elements, and electron-positron annihilation freeze-out. }
    \end{figure}

    The numerical solution of the electron/positron SNSI contributing to the neutrino mass is depicted in Fig. \ref{fig:mass_correction}.
    At high temperatures, both contributions to the effective mass are the same.
    Below the electron mass threshold, the contribution decays exponentially as the Universe cools down.
    But, when the electron-positron annihilation ends, only electrons remain in the background. 
    However, at temperatures much smaller than $\mathcal{O}$MeV, the neutrino mass correction contribution induced by leptons becomes negligible.
    
    Unlike some terrestrial and astrophysical scenarios, here we also need to consider the background composed of relic neutrinos.
    In this self-interacting case, $\Delta m$ would have another unknown parameter, the bare neutrino mass $m_{\nu}$.
     Notice that, in order to have a $\Delta m$ of the same order of magnitude than the one induced by charged leptons, $G_{{\rm S}}$ needs to be roughly $m_{{\rm e}}/m_{\nu}$ times larger than $G_{{\rm eff}}$, see equation (\ref{eq:deltam}). 
    As the BBN epoch occurs at temperatures much larger than the bare neutrino mass scales, the mass correction does not drop exponentially with the temperature as it occurs with the electron/positron SNSI.
    By definition, by constraining $G_{{\rm S}}$, we would be able to find a mediator mass-dependent $g_{\nu}$-bound.
    
    By oscillation experiments, we know that at least two neutrinos are massive.
    Hereafter, we shall take a conservative value for bare neutrino masses, being one-third of the minimum sum of neutrino masses in the normal hierarchy, $(\sum m_{\nu})_{{\rm min}} \sim 0.059$ eV \cite{Loureiro2019}, assuming an almost degenerate scenario of active neutrinos.
    Given this, we take $m_{\nu}=0.0195$ eV and assume all three neutrino parameters are universal.

    \section{\label{sec:cosmo_imp} Cosmological implications}

    In the previous section, we have described how the neutrino $m_{\rm eff}$ would be affected by CFS with charged leptons and neutrinos at high temperatures.
    We now focus on the implementation and implications of neutrino SNSI in the Early Universe.
    In particular, in this section we compute $N_{\rm eff}$ as a function of the SNSI parameters.
    
    The particles in the plasma are in local thermal equilibrium when their interaction rate is larger than the rate of the expansion of the Universe, $\Gamma \gg H(T_{\gamma})$.
    The Universe at high temperatures ($T_{\gamma} \sim \mathcal{O}$MeV) is dominated by radiation and the density of any heavy particle, $m \gtrsim T_{\gamma}$, gets suppressed.
    A large neutrino $m_{\rm eff}$ will diminish its production by weak interactions and ultimately the Universe will have less radiation than expected.
    Therefore, by weighting the effect of $m_{\rm eff}$ on $N_{\rm eff}$, we will estimate the permitted values of the SNSI parameters.
    
    As we stated in the previous section, the mass correction diagram of Fig. \ref{fig:tadpole} is describing a CFS  that implies no transfer of energy and momentum with the plasma.
    Therefore, a priory, the neutrino thermal evolution should remain unchanged.
    Nonetheless, we carefully explore whether the $m_{\rm eff}$ is capable of changing the thermal evolution of neutrinos.

    In this scenario, the weak interaction is the one responsible for keeping neutrinos in thermal equilibrium with the plasma.
    In equilibrium, the neutrino energy and number density, for one flavor, are given by \cite{mukhanov2005physical}
    \begin{eqnarray}\label{eq:energy_density}
        \rho_{\nu}(G_{{\rm eff}},G_{{\rm S}};T_{\nu},T_{\gamma})=\frac{T_{\nu}^4}{\pi^2}\int_{\alpha}^{\infty}\frac{dx\  x^2 \sqrt{x^2-\alpha^2}}{e^x+1},\\
        n_{\nu}(G_{{\rm eff}},G_{{\rm S}};T_{\nu},T_{\gamma})=\frac{T_{\nu}^3}{\pi^2}\int_{\alpha}^{\infty}\frac{dx\  x \sqrt{x^2-\alpha^2}}{e^x+1},\nonumber
    \end{eqnarray}
    where $x=E_{\nu}/T_{\nu}$ and $\alpha = m_{{\rm eff}}/T_{\nu}$. The effective mass
    $m_{{\rm eff}}$ encodes all the new physics, as given in equation \eqref{eq:meff}.
    Notice that the neutrino density gets suppressed with a larger $m_{{\rm eff}}$.
    As the Universe cools down, the effective neutrino mass drops significantly, this permits the neutrino density to approach and possibly recover its standard value.
    However, after neutrino decoupling, is not possible to produce abundantly new neutrinos to reach their standard density.
    Thus, establishing the neutrino decoupling temperature is important to compute the final neutrino density to a good approximation.
    
    The neutrino thermal mass, if relevant, would increase the temperature at which neutrinos decouple.
    We compare the interaction rate of electron-neutrino scattering, which is the responsible to keep neutrinos in equilibrium, with the expansion rate.
    In the standard theory, we have $\Gamma_{ew} \propto (1-m_e^2/T^2)^2\ T^5$, while the expansion rate is proportional to the energy density $H(T)\propto \sqrt{\rho}$. In the radiation dominated epoch  $\rho\propto T^4$, and thus $H\propto T^2$.
    The SNSI effect diminishes both the interaction and the expansion rates.
    Although, we have found that the dominant effect comes from the interaction rate depletion.
    We estimate the ratio of the cross-section to the SM one to be \cite{greiner1996gauge}
    \begin{equation}\label{cross_section_ratio}
    \frac{\sigma}{\sigma_{\rm SM}}=\frac{2\sqrt{A_+ A_-}}{3\left(1-m_e^2/T^2\right)^2}[A_+A_-+1 - \frac{A_++ A_-}{4}-B_-^2],
    \end{equation}
    where $A_+=1-((m_e+m_{\rm eff})/T)^2$, $A_-=1-((m_e-m_{\rm eff})/T)^2$, $B_-=(m_e^2-m_{\rm eff}^2)/T^2$, and the condition $m_e+ m_{\rm eff}<T$, which is true for the permitted parameter region we will present in the next section. Note that in the limit $m_{\rm eff}\rightarrow 0$ we have $\sigma/ \sigma_{\rm SM}=1$.
    
\begin{figure}
    \includegraphics[width=0.47\textwidth]{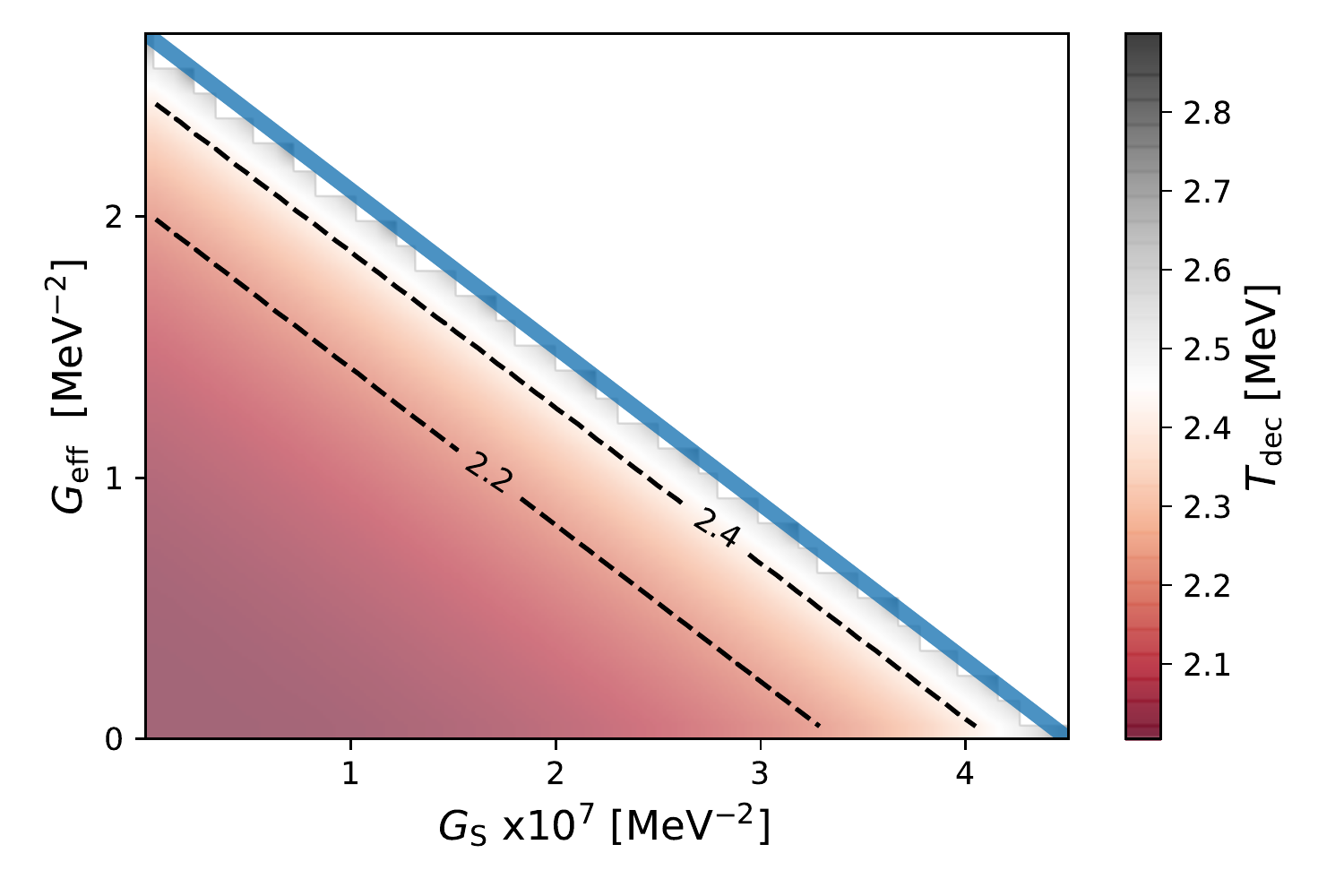}
    
    \caption{\label{fig:decoupling_temp} Neutrino decoupling temperature as a function of the SNSI parameters. The blue line denotes the threshold where the neutrino effective mass becomes too large spoiling its standard thermalization.}
    \end{figure}

    We solve numerically the equation $\Gamma=H$ for each pair ($G_{\rm S}$,$G_{\rm eff}$) to find the decoupling temperature due to SNSI.
    We model the interaction rate as $\Gamma=\left<\sigma v \right>n_{e}=\xi \sigma n_e$, where $\xi$ encodes our ignorance about the thermal average, $n_{e}=3\zeta(3)T^3/(2\pi^2)$ is the electron/positron density, and
    \begin{equation}
        \sigma=\frac{2}{3\pi}G_{\rm F}^2T^2\sqrt{A_+A_-} [A_+A_-+1 -\frac{A_++ A_-}{4}-B_-^2],
    \end{equation}
     where $G_{\rm F}\sim1.166 \times 10^{-11}$ MeV$^{-2}$ is the Fermi constant.
    We set the value of $\xi\sim 6.5$ to match the most conservative value for the standard case $T_{\rm dec\ std}=2$ MeV.
    Furthermore, we assume the value of $\xi$ does not change due to SNSI.
    Interestingly, we find a region of the parameter space ($Gs>4.6\times 10^7$ MeV$^{-2}$, $G_{\rm eff}>2.8$ MeV$^{-2}$, or other combinations) where $\Gamma$ is always smaller than $H$.
    This region exhibits an exotic behavior that suggests that  for the very large thermal mass corrections, neutrinos may not get into thermal equilibrium with the radiation plasma. Avoiding such a non-physical scenario imposes a natural bound on the thermal mass and thus to the couplings.
    In Figure \ref{fig:decoupling_temp} we show the numerical results for $T_{\rm dec}$ in the SNSI parameter grid, there the parameter region where the thermal mass surpasses acceptable values had been excluded.

    The neutrino density freezes out at their decoupling temperature, and no significant amount of neutrinos gets produced after that.
    This is because weak interactions would be able to produce only a small percentage of the total neutrino density.
    Lastly, notice that assuming this late instantaneous neutrino decoupling is the most conservative approach.
    However, we expect almost the same final neutrino density than using a more complex model for neutrino decoupling.
    Since, in the standard case, neutrinos decouple the earliest at $\sim 3$ MeV.
    
    \begin{figure*}
    \includegraphics[width=0.47\textwidth]{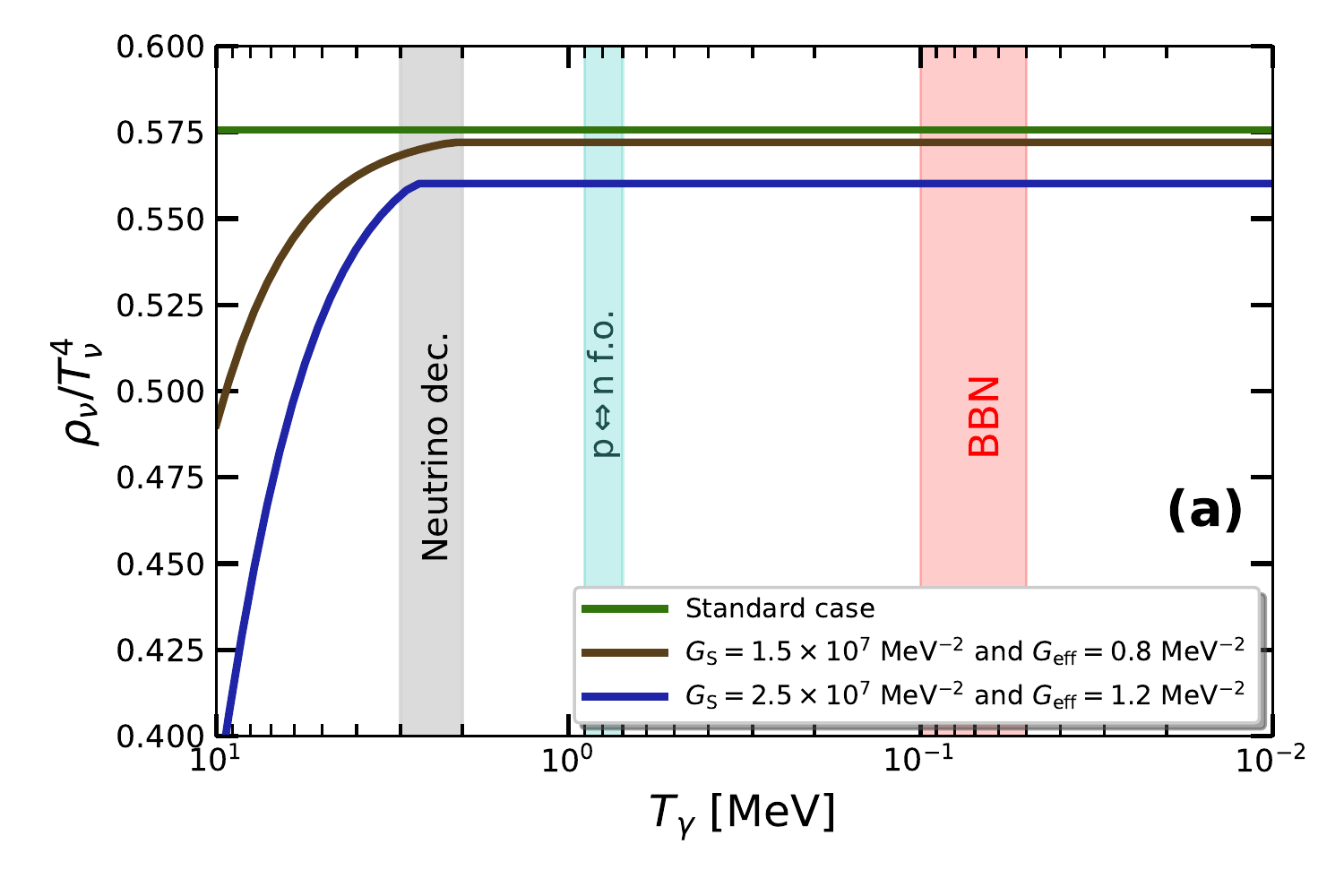}
    \includegraphics[width=0.47\textwidth]{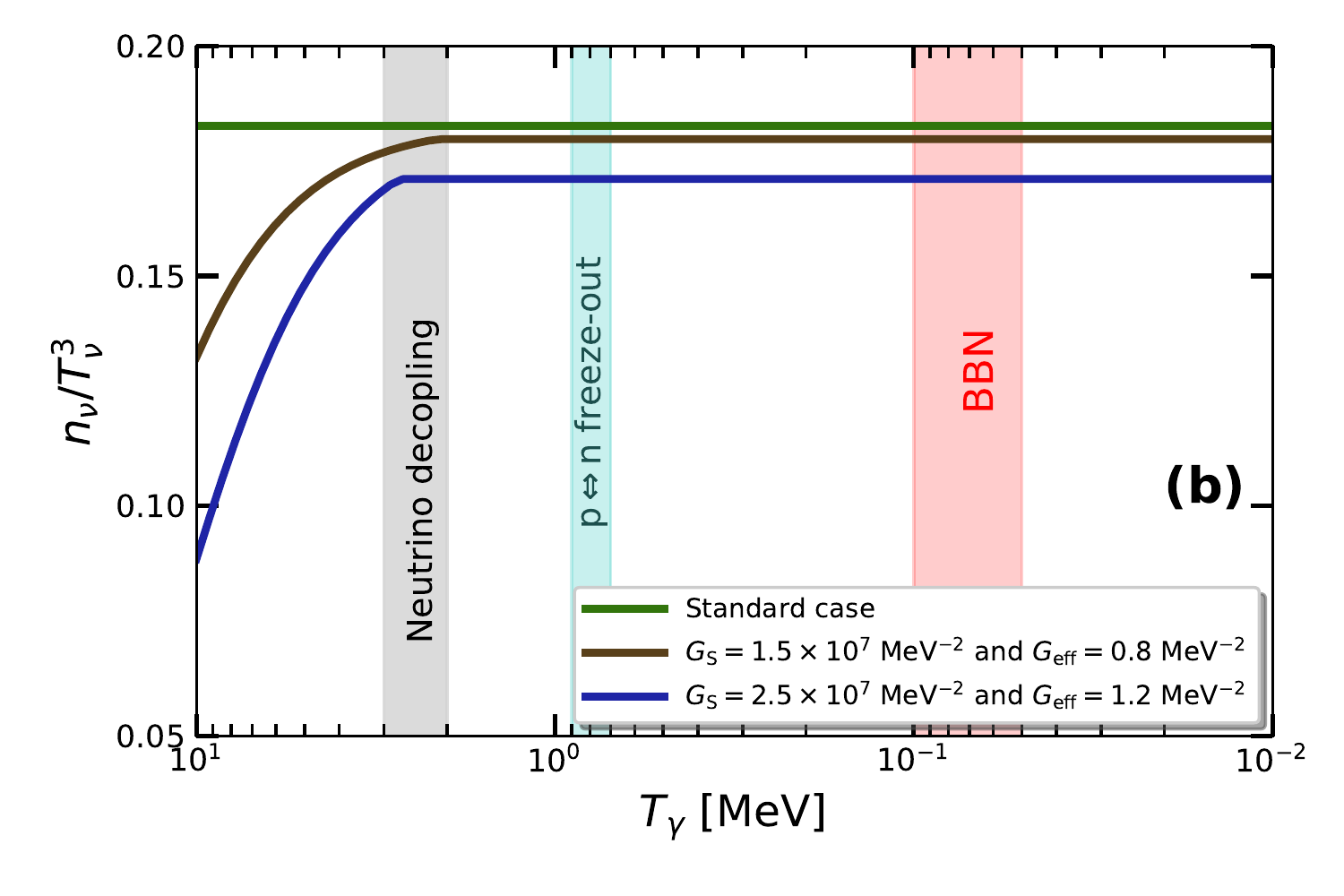}
    \includegraphics[width=0.47\textwidth]{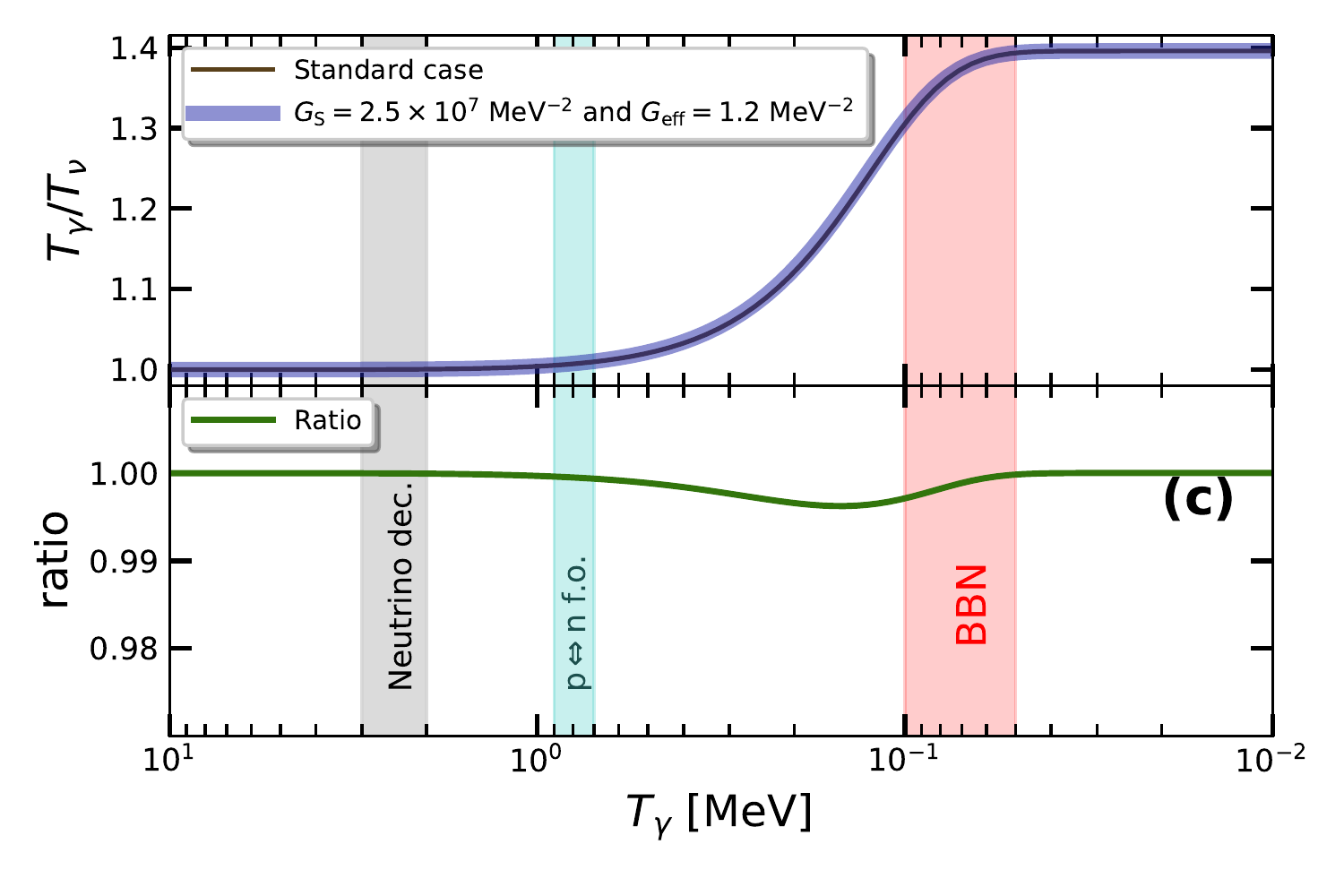}
    
    \caption{\label{fig:density} Neutrino density and temperature evolution for different values of the SNSI effective couplings. Key cosmological events highlighted: Neutrino decoupling, proton to neutron freeze-out, and synthesis of light elements. (a)Neutrino energy density evolution. (b)Neutrino number density evolution. (c)Evolution of neutrino temperature respect to the photon temperature.}
    \end{figure*}

    We move on to model the neutrino density.
    The energy density becomes a piece-wise function, where the neutrino density freezes out at the threshold  $T_{\gamma}=T_{\rm dec}$, 
    \begin{align}
    \rho_{\nu} = \left\{ \begin{array}{cc} 
                    \rho_{\nu}(G_{{\rm eff}},G_{{\rm S}};T_{\nu},T_{\gamma}) &  T_{\gamma} > T_{\rm dec}\\
                    \left(\frac{T_{\nu}}{T_{\rm dec}}  \right)^4\rho_{\nu}(G_{{\rm eff}},G_{{\rm S}}; T_{\rm dec},T_{\rm dec}) &  T_{\gamma}\le T_{\rm dec} \\
                    \end{array} \right. 
    \end{align}
    where $\rho_{\nu}(G_{{\rm eff}},G_{{\rm S}};T_{\nu},T_{\gamma})$ is the thermal density described in \eqref{eq:energy_density}.
    After decoupling, the neutrino density falls due to the adiabatic expansion of the Universe. We sketch this in Figs. \ref{fig:density} (a) \& (b) for different values of the SNSI effective parameters.
    
    We also track the neutrino temperature evolution after decoupling.
    We employ a modified version of the public available code \textsc{nudec\_bsm} \cite{Escudero2020,Escudero2019_Neutrino_decoupling_BSM}.
    This code solves for the ratio of the neutrino and photon temperature in a much simpler approximation than state-of-the-art codes \cite{de_Salas_2016neff}.
    Unlike other precise computations of $N_{{\rm eff}}$, this code does not include neutrino oscillations, yet, it computes a pretty robust value of $N_{{\rm eff}}=3.045$ in the SM case.
    In Fig \ref{fig:density} (c) we depicted the evolution of the temperatures for the standard case and a scenario denoted as large $G_{{\rm eff}}$ and $G_{{\rm S}}$ values.
    We observe that the evolution of temperatures differs only within the numerical error values.
    
\begin{figure}
    \centering
    \includegraphics[width=0.47\textwidth]{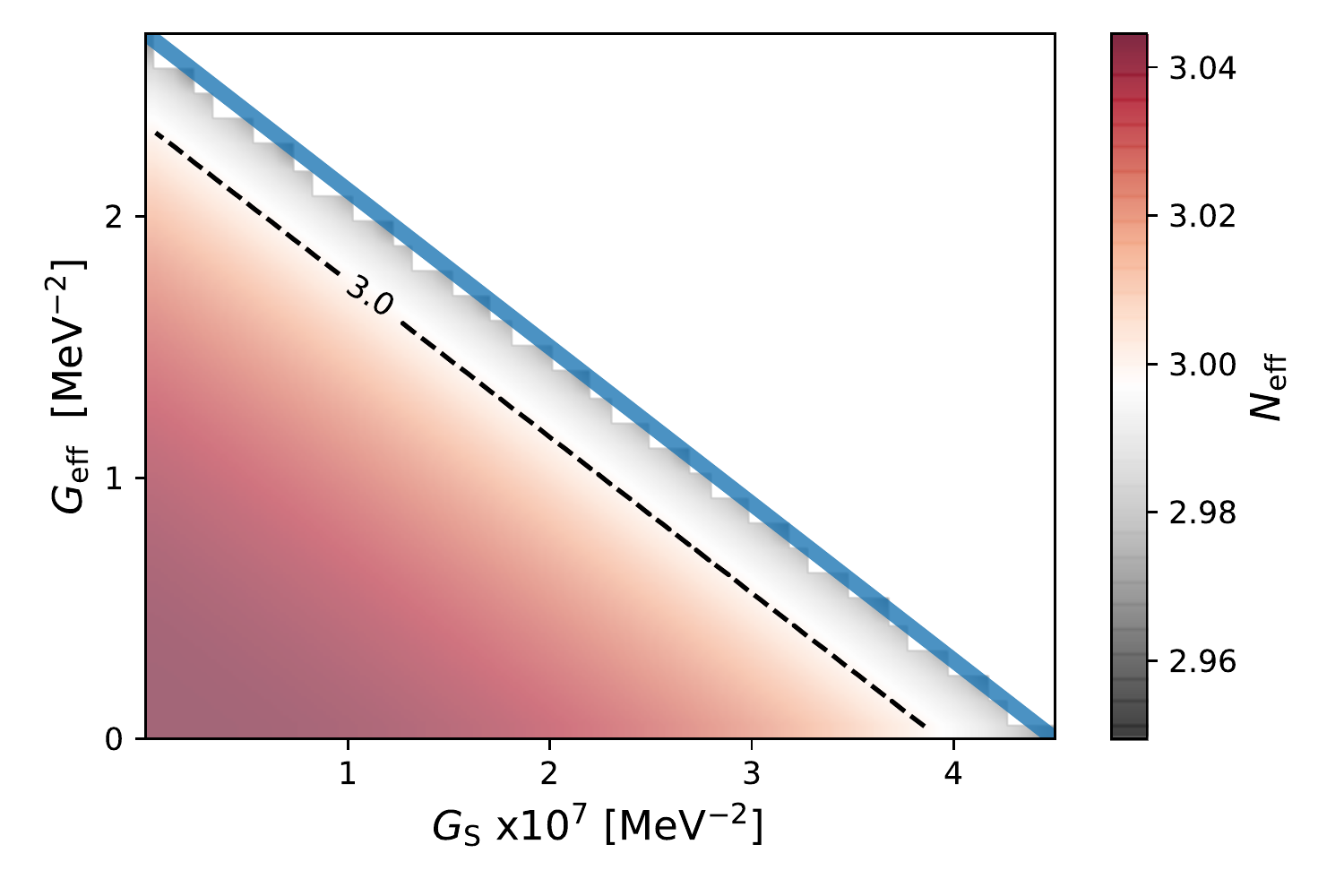}
    
    \caption{\label{fig:grid_vs_neff} $N_{{\rm eff}}$ as a function of the effective couplings $G_{{\rm S}}$ and $G_{{\rm eff}}$. The blue line denotes the threshold beyond which the effective mass spoils neutrino standard thermal history.}
    \end{figure}

    We proceed to numerically compute the effective number of relativistic species $N_{{\rm eff}}$.
    For this purpose, we again employ the code \textsc{nudec\_bsm}.
    We observe that there is a direct relation between the SNSI parameters and $N_{{\rm eff}}$, this is given by
    \begin{equation}\label{eq:neff}
        N_{{\rm eff}}(G_{{\rm eff}},G_{{\rm S}}) \equiv \frac{8}{7}\left(\frac{11}{4}\right)^{4/3} \frac{3\rho_{\nu}(G_{{\rm S}},G_{{\rm eff}})}{\rho_{\gamma}} \hspace{2mm} {\rm for} \hspace{1mm} T_{\gamma}\ll m_e,
    \end{equation}
    where  $\rho_{\gamma}=(2\pi^2/30) T_{\gamma}^4 $ is the photon density and we have assumed a full degeneration of neutrino parameters.
    In Fig. \ref{fig:grid_vs_neff} we illustrate the change on $N_{{\rm eff}}$ as a function of the effective couplings $G_{{\rm S}}$ and $G_{{\rm eff}}$, where we  used  $T_{\nu}/T_{\gamma}\sim 0.7164$ as obtained from the code \textsc{nudec\_bsm}. Notice that in the limit $G_{\rm S}, G_{\rm eff}\rightarrow 0$, we recover $N_{\rm eff}\simeq 3.04$.
    We can observe a strong positive correlation between the SNSI parameters since they both produce the same effect.
    In the next section, we will constraint these parameters with BBN physics.
    
    Lastly, notice that we are neglecting first order SNSI processes, such as three-level scattering.
    Thus, we assume that they do not contribute to any change in the neutrino temperature, production, and decoupling.
    Besides, we assume that the scalar mediator is out-of-equilibrium with the plasma, so it does not acquire a thermal mass, nor is being thermally produced within the plasma at a significant number.
    Finally, we assume that neutrino decays into the scalar are irrelevant, hence, the scalar density is insignificant.

    Dimensional arguments permit us to explore the validity of these approximations.
    Our region of interest lies in the small scalar mass regime $m_{\phi}\ll$ keV.
    In this case, the SNSI cross-section of processes such as $e^- \nu \rightarrow e^- \nu $ would be $\sigma_{{\rm SNSI}} \approx g_e^2g_{\nu}^2/T^2 $.
    While, the SM cross-section is given by $\sigma_{{\rm SM}} \approx \alpha^2T^2/M_{w}^4$, where $\alpha \sim 1/137$ is the fine-structure constant and $M_{w}\sim 80$ GeV is the W Boson mass. 
    Comparing both cross-sections, we observed that the condition $g_eg_{\nu}< \alpha T^2/M_{w}^2 $, for $T\sim 1$ MeV implies that $g_eg_{\nu}\lesssim10^{-12}$.
    Similarly, the scalar would be prevented from reaching thermal equilibrium as long as the condition $g_{\nu}^2<\alpha T^2/M_{w}^2$ is satisfied.
    Lastly, the scalar would not significantly contribute to $N_{\rm eff}$, provided the condition $g_{e}<g_{\nu}\lesssim 10^{-5}$ for  $m_{\phi}\ll$keV \cite{Huang2018} is not violated.
    As we will argue along section \ref{sec:comparison}, these conditions would be satisfied in the ultralight scalar regime (see Figs. \ref{fig:mphi_vs_gnu} and \ref{fig:mphi_vs_ge}).

    \section{ \label{sec:BBN} BBN constraints}
    
    \begin{figure*}
    \centering
    \includegraphics[width=0.48\textwidth]{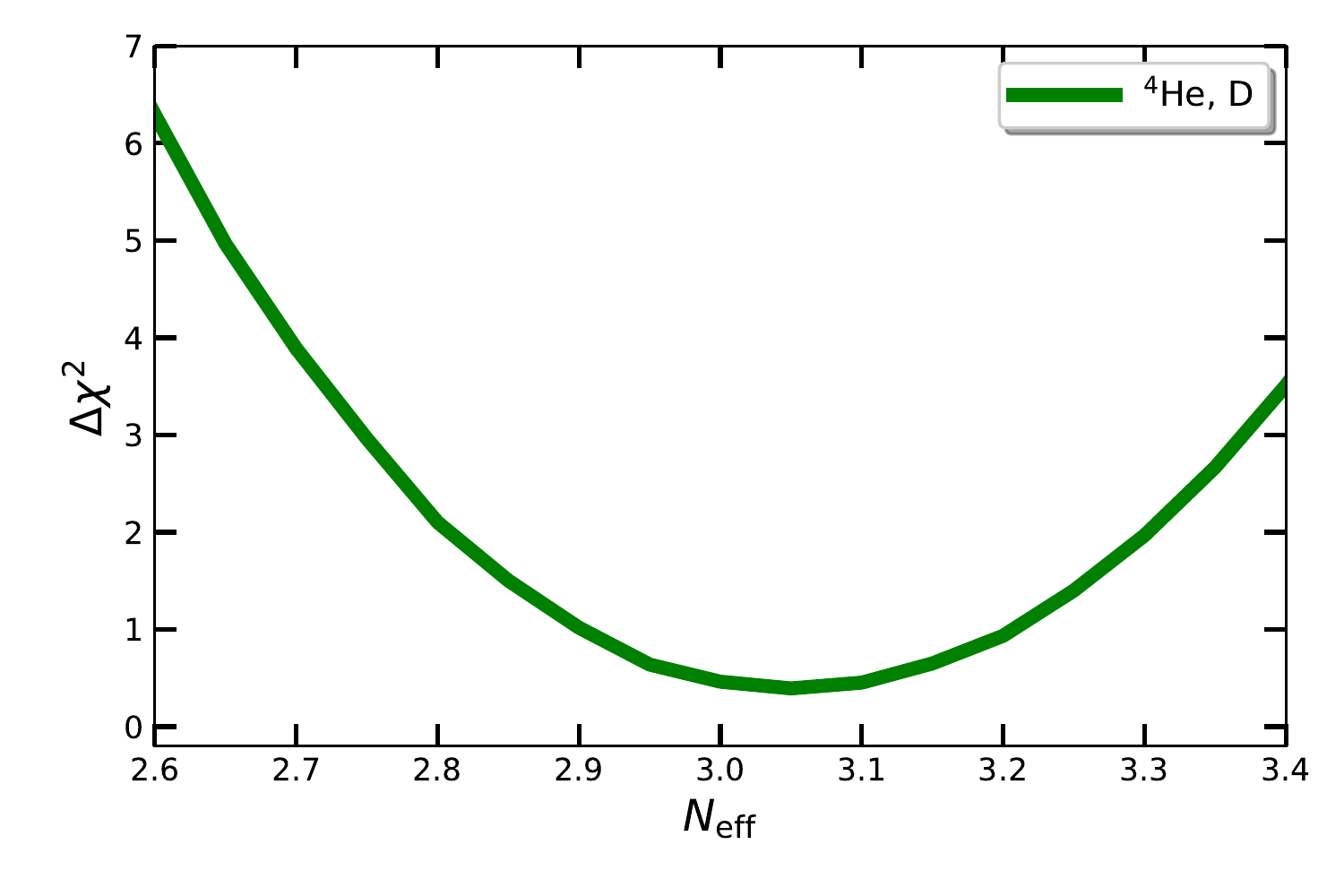}
    \includegraphics[width=0.48\textwidth]{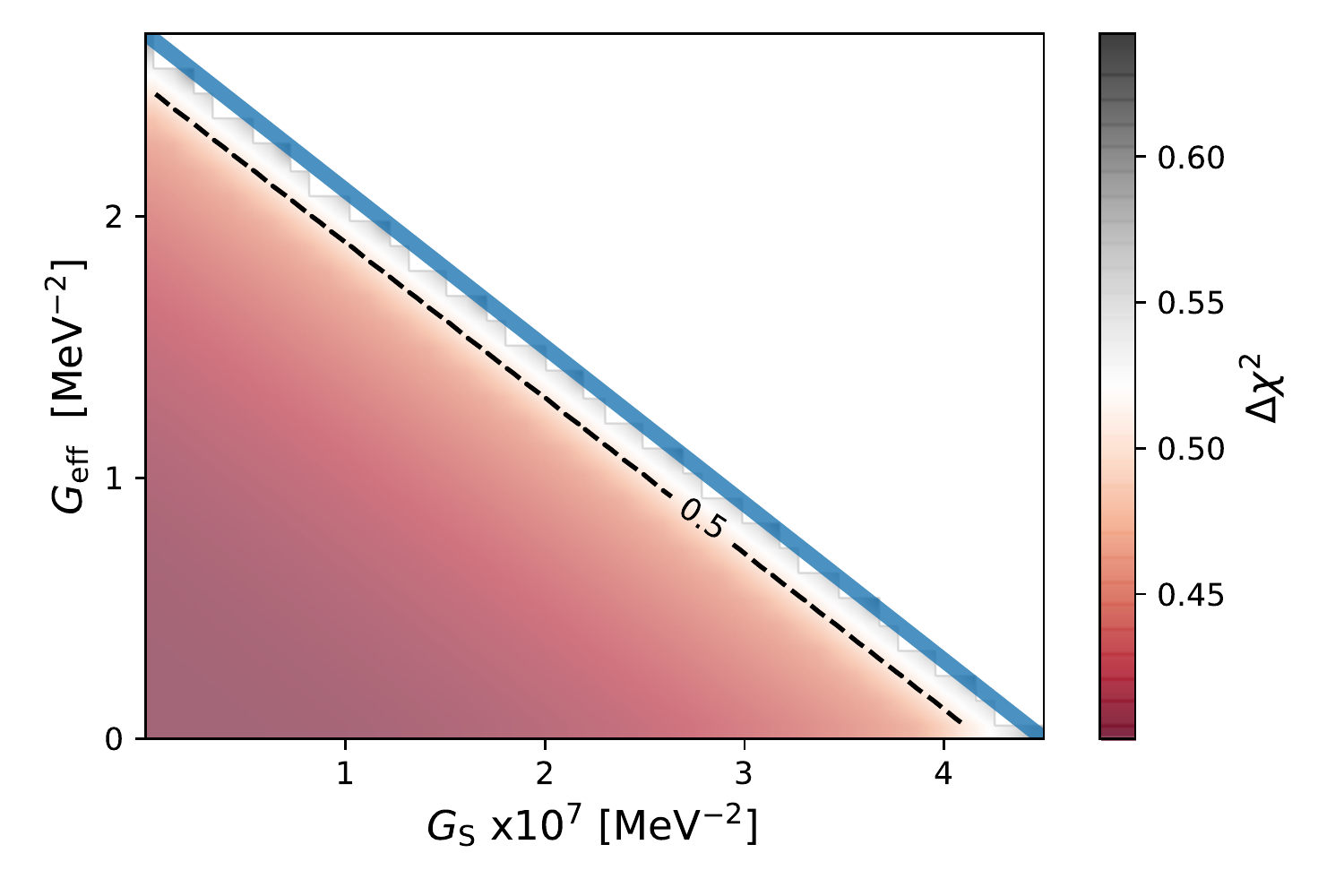}
    \caption{\label{fig:delta_chi2} $\chi^2$ as a function of the model parameters with $\eta=6.11\times 10^{-10}$. \textit{Left panel:} $\Delta \chi^2$ as a function of the effective number of relativistic species $N_{{\rm eff}}$. Measurements of helium-4 and deuterium. \textit{Right panel:} $\Delta \chi^2$ as a function of the effective couplings $G_{{\rm S}}$ and $G_{{\rm eff}}$. As before, the blue line denotes the threshold above which neutrino standard thermalization gets spoiled.}
    \end{figure*}

    In this section, we present the bounds on the effective SNSI parameters $G_{{\rm S}}$ and $G_{{\rm eff}}$ by BBN theory and observations of light element abundances.
    
    BBN is one of the cornerstones of modern cosmology and the Big Bang Theory. With the interplay of standard nuclear and particle physics and the standard cosmological model, it describes with great accuracy the synthesis of the lighter nuclei during the very first seconds of cosmic time (For a review see~\cite{Cyburt2016}). Despite some uncertainties on the predictions for \ce{^7Li}, which may have a diversity of possible sources \cite{Fields2011,Starrfield2020}, it predicts the observed relative abundances of \ce{H}, \ce{D}, \ce{^3He}, \ce{^4He} and \ce{^7Li} as a function of a single  parameter, the baryon-to-photon ratio, $\eta=n_b/n_\gamma$, or equivalently, the present baryon density $\Omega_b h^2$, which determines the end of the deuterium bottleneck, and therefore, the production rate of heavier nuclei.
    
    Aside from the initial condition on $\eta$, which is thought to be associated with an earlier baryogenesis process, for which the SM seems to have not a satisfactory explanation, BBN success is based on well-known physics, which leaves little space for new or exotic physics. This feature is precisely what makes BBN a useful probe for any nonstandard physics that may modify the cosmological evolution during those early times. In particular, any physics that could change the expansion rate during BBN \cite{Peimbert2016,Lague2020,Foot1997}. As the effective neutrino mass that we are discussing changes $N_{{\rm eff}}$, it does affect the amount of radiation during that epoch, and so BBN should be sensitive to it. We will focus on this in what follows.
    
    To a good approximation, when the deuterium bottleneck breaks up, most of the neutrons present in the primordial Universe are synthesized in \ce{^4He}. Other elements are then produced at much smaller amounts, with a rate of about $10^{-5}$ for \ce{D} and \ce{^3He} and $10^{-10}$ for \ce{^7Li} per proton. \ce{^4He} mass fraction is well approximated as
    \begin{equation}
        Y_p \approx \frac{2(n/p)}{1+(n/p)},
    \end{equation}
    where the neutron to proton ratio, $(n/p)$, at BBN is determined by the output ratio at weak interactions freeze out, when the weak interactions rate per baryon  $\Gamma_{ew} \approx \alpha^2 T^5/M_w^4$ becomes smaller than Hubble expansion, and by neutron number depletion due to $\beta$ decay. In the standard cosmological model, at temperatures well above the neutron - proton mass difference, $ \Delta m = m_n-m_p=1.239$ MeV,  neutrons and protons are in chemical equilibrium. Below that temperature electron neutrino capture process, $n\nu\rightarrow pe$, starts favoring protons. Since $\eta$ is small, this process does not  sensitively alter lepton population and, thus, the corresponding Boltzmann equation is written as
    \begin{equation}\label{XnBoltzmann}
        \frac{dX_n}{dt} = \lambda_{np}\left[(1-X_{n})e^{-\Delta m/T}- X_{n}\right]~,
    \end{equation}
    where $Y_p = 2 X_n$ and $\lambda_{np}=n_\nu^{(0)}\langle\sigma v\rangle$. Note that the last is mostly independent of $n_\nu^{(0)}$, but sensitive to neutrino spectrum, since $\langle\sigma v\rangle={\cal I}/(n_\nu^{(0)} n_n^{(0)})$, with $\cal I$ an integral over all particles momentum space of the differential cross section weighted by the Boltzmann factor $e^{-(E_\nu +E_n)/T}$.
    At freeze out temperature, $T_\star\sim 0.8~$MeV,  $(n/p)_\star\simeq e^{-\Delta m~T_\star}\sim 1/5$ and thus one estimates $(n/p)_{\rm BBN} \simeq 1/7$~(for a theoretical calculation of this see for instance \cite{Bernstein1989,Mukhanov2004}).
    
    The key feature for our present analysis resides in the fact that setting $N_{\rm eff}$ as a free parameter compromises the expansion rate during the radiation dominated epoch. A  smaller (larger) value of $N_{\rm eff}$ than the one computed in the standard case, reduces (increases) expansion rate and lowers (raises) weak interactions decoupling temperature. Even if the change is mild, due to Boltzmann suppression, a smaller (larger) $T_\star$ implies a lower (higher) $(n/p)$ and thus a smaller (higher) $Y_p$.
    Notice that there is a competing effect when an excess (deficit) of neutrinos over equilibrium spectrum exists, since it increases (decreases) weak rates, implying a smaller (larger) $Y_p$~\cite{Dolgov2002}. However, as effective neutrino thermal mass mainly affects the low energy part of the spectrum, where the differential cross section quickly dies down,  last effect is expected to be less relevant against  varying $N_{\rm eff}$.

     CMB is sensitive to both $\eta$ and $Y_p$ and as a matter of fact, Planck data alone provides a determination of $Y_p$~\cite{Aghanim2018Planck}.
     Although $Y_p$ is not sensitive to the baryon-to-photon ratio, as we mentioned earlier, $\eta$ is an important initial condition for BBN and the production of other light elements.
     Here, we keep our analysis consistent with CMB using a prior for $\eta$ consistent with the permitted region at 1-sigma by Planck data $\Omega_b h^2=0.0224\pm 0.0001$, or equivalently, $\eta=6.11\pm 0.03 \times 10^{-10}$.
    
    In order to constraint the SNSI effective parameters, we use the observations of primordial deuterium and helium abundances.
    For our porpuse, we employ a modified version of the public code \textsc{alterbbn}  \cite{Arbey2012,AlterBBNv2}, where, hereafter, we use a neutron lifetime $\tau_n=880.2$ s.
    We use a $\chi^2$-analysis with
    \begin{equation}
        \chi^2 = \sum \frac{(R_{\rm{SNSI}} - R_{\rm{obs}})^2}{\sigma^2} \,,
    \end{equation}
    where $R_{\rm{obs}}$ and $R_{\rm{SNSI}}$ are respectively the observed and theoretical nucleon fractions and $\sigma$ its observational error \cite{tanabashi2018review}. The sum is over the two measurements of helium and deuterium fractions, $Y_p=0.245 \pm 0.003$ and ${\rm D/H}=(2.569\pm 0.027)\times 10^{-5}$.
    In Fig. \ref{fig:delta_chi2} (LHS) we present the fluctuation of the $\chi^2$ function $\Delta \chi^2 =\chi^2- \chi^2_{{\rm min}}$ as function of the parameter $N_{{\rm eff}}$. 
    Notice that the $\chi^2$ is not symmetric and gets steeper for small values of $N_{{\rm eff}}$.
    Therefore, this lets us set stringent constraints to the SNSI parameters.
    
    As we mention earlier, we obtain the constraints on the SNSI effective parameters by taking the advantage of the direct relation between $N_{{\rm eff}}$ and a pair $G_{{\rm S}}$ and $G_{{\rm eff}}$ (see Fig. \ref{fig:grid_vs_neff}).
    In Fig. \ref{fig:delta_chi2} (RHS) we present the deviation of $\chi^2$ from its minimum value as a function of the SNSI parameters.
    Notice that, we have two strongly correlated parameters, thus, we employ a statistical procedure in which we can find a robust bound for each parameter.
    
    In order to find the bounds for $G_{\rm{eff}}$ and $G_{\rm S}$ we use the posterior distribution $P(\theta_1, \theta_2,\eta)\propto e^{-\chi^2/2}$ where $\theta_1$ is the parameter that we are analyzing, either $G_{\rm{eff}}$ or $G_{\rm S}$. We marginalize over the second parameter $\theta_2$ and the baryon-to-photon ratio $\eta$ to obtain the single parameter posterior distribution $P^{(1)}(\theta_1)\propto \int \int P(\theta_1,\theta_2) d\theta_2 d\eta$. Then we follow \cite{Hamann_2007} to find the credible intervals for our situation. In our case the minimum of the credible intervals coincide with the physical bound of the parameters $G_{{\rm S}}$ and $G_{{\rm eff}}$ which are constrained to be positive. The $100\gamma \%$ credible region will be defined as
    \begin{equation}
        \int_{0}^{\theta_{1\rm{bound}}} d\theta_1 P^{(1)}(\theta_1) = \gamma_1 \,.
    \end{equation}
    Using a finite grid the $\gamma$-value for the parameter $\theta_1$ is
    \begin{equation}
         \gamma_1=\frac{1}{N}\sum_{i=0}^{i_{\rm bound}}\sum_{j,k}{\rm e}^{-\chi^2_{ijk}/2}\  \delta \theta_{1 i} \delta \theta_{2 j}\delta \eta_{j}~,
    \end{equation}
    where $N=\sum_{i,j,k}{\rm e}^{-\chi^2_{ijk}/2} \delta \theta_{1 i} \delta \theta_{2 j}\delta \eta_{j}$, $\delta \theta$ is a small constant finite difference in the parameter sampling, and $\delta \eta$ is similarly defined.

    We obtain the parameter constraints at 68\% CL by finding the $\theta$-values that make $\gamma_1=\gamma_2=0.68$. Our final marginalized bounds are 
    \begin{eqnarray}\label{Gbounds}
    G_{{\rm eff }}< 1.2\ {\rm MeV}^{-2} \ \ \ (68\% \ {\rm CL}) .\\
    G_{{\rm S }}< 2.0 \times 10^{7}\ {\rm MeV}^{-2} \ \ \ (68\% \ {\rm CL}). \nonumber 
    \end{eqnarray}
  Notice that, although parameter marginalization is the appropriate statistical procedure to obtain robust constraints, the effective SNSI parameters are strongly positively correlated and the degeneracy cannot be entirely broken up.
    
    In the next section, we will disentangle the model parameters (the couplings and the mediator mass) by using our constraints in the effective SNSI parameters and present new bounds on the neutrino-scalar coupling.

    \section{\label{sec:comparison} Parameter space comparison}
    
    \begin{figure}
    \centering
    \includegraphics[width=0.48\textwidth]{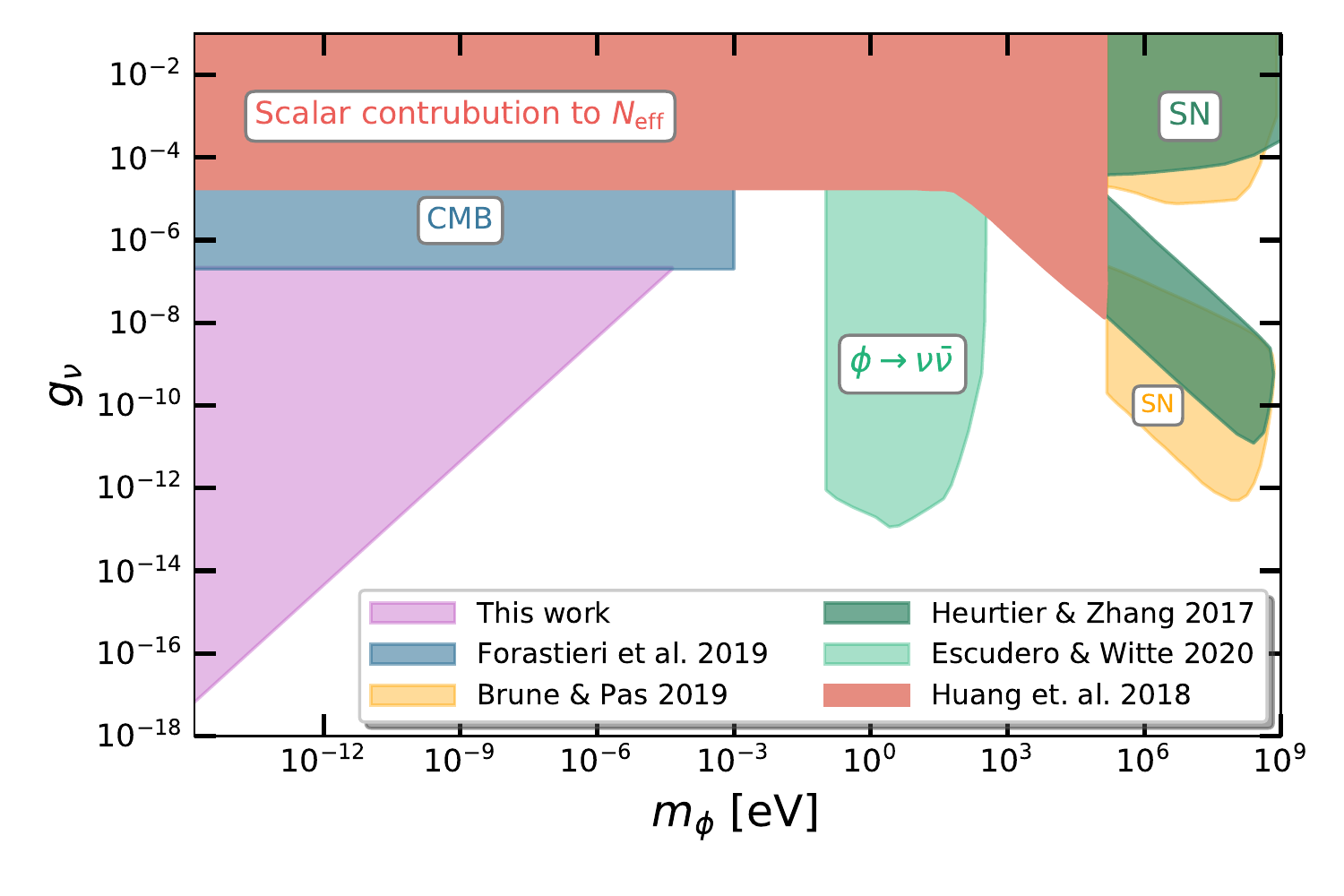}
    \caption{\label{fig:mphi_vs_gnu} Neutrino-scalar coupling constraints. The purple area represents the region excluded by our bound on $G_{{\rm S}}<2.0 \times 10^{7}$ MeV$^{-2}$~. The blue area represents the CMB excluded region for scalar mediated self-interacting neutrinos~\cite{Forastieri2019}.
    The turquoise area represent the CMB constraints by scalar decays into neutrinos \cite{Escudero2020_CMB_search}.
    The red area represents the excluded region for a scalar contributing to $N_{{\rm eff}}$ during BBN \cite{Huang2018}. Green and orange areas portray the region excluded by supernovae \cite{Heurtier2017,Brune2019}. }
    \end{figure}

    We have discussed and computed the bounds on the effective parameters of SNSI.
    Here we translate the bounds into the mass-coupling parameter space and compare our constraints with others from terrestrial experiments as well as astrophysical/cosmological observations.
    
    First, we should discuss one important limit in the cosmological approach.
    As stated by \cite{Babu2020scalar} the scalar mass has a lower bound imposed by the size of the Universe at the relevant epochs. This is because the De Broglie wavelength of the particle, $l\propto m_{\phi}^{-1}$, cannot be larger than the size of the Universe. Otherwise, it will escape the Hubble horizon.
    For our considerations, we establish the scalar mass lower bound from Hubble radius at 2 MeV, $H^{-1}(2{\rm MeV})$.
    Thus, our results are only valid for $m_{\phi}\gtrsim1.5 \times 10^{-15}$ eV.
    Interestingly, notice that the size of the Hubble horizon at those epochs is smaller than the size of the Sun.
     
    Here, we present a new stringent bound on the scalar-neutrino coupling $g_{\nu}$, which is particularly robust for ultralight scalar masses.
    Note that the  bound on $G_{{\rm S}}$ [from Eq.~(\ref{Gbounds})] permits us to find a mass-dependent bound on $g_{\nu}$, that goes as
    \begin{equation}\label{eq:gnu}
    g_{\nu}< 4.5\times 10^{-3} \left(\frac{m_{\phi}}{{\rm eV}}\right)\ \ \ (68\% \ {\rm CL}).
    \end{equation}
    This new bound restricts a large new region in the parameter space ($m_{\phi}$,$g_{\nu}$) for masses $1.5 \times 10^{-15} {\rm eV}\lesssim m_{\phi}\lesssim 4.5 \times 10^{-5}{\rm eV}$, where the upper value comes from $g_{\nu}<2\times 10^{-7}$ derived by the authors in \cite{Forastieri2019}.

    In the literature, we spot that there have been extensive efforts to impose bounds on the neutrino-scalar coupling.
    For instance, it has been constrained by coherent elastic neutrino-nucleus scattering (CE$\nu$NS) and by scalar emission in neutrinoless double beta decay experiments
    \cite{Blinov2019,Pasquini2016,farzan2018,Brune2019}.
    Interestingly, a neutrino-scalar coupling around $\sim 10^{-6}$ could explain the recent anomalous spectral excess at the \textsc{XENON1T}  experiment \cite{Khan2020Xenon1t,xenon1t2020}.
    
    Astrophysical and cosmological observations set the strongest constraints on the neutrino-scalar coupling. 
    A neutrino flavor-dependent scalar interaction is responsible for several non-observed effects in supernovas (SN).
    Such effects include a loss of SN luminosity, loss of leptons in the supernova core (deleptonization), and trapping of neutrinos by dispersion with a (pseudo)scalar.
    In Fig. \ref{fig:mphi_vs_gnu}, we depict the strongest SN bounds, corresponding to a (pseudo)scalar coupled to electron neutrinos $|g_{{\rm ee}}|$ \cite{Heurtier2017,Brune2019}.
    
    We also revisit a couple of cosmological bounds. 
    The bound imposed by \cite{Huang2018} confronts the positive contribution of a light scalar particle to $N_{{\rm eff}}$, namely $\Delta N_{{\rm eff}}$, with BBN physics.
    On the other hand, in \cite{Forastieri2015,Forastieri2019}, the authors studied the observable effects on the CMB caused by neutrino self-interactions mediated by a very light scalar particle $m_{\phi}\ll T_{\nu}$.
    As neutrinos become collisional again at small temperatures, this approximation holds for $m_{\phi}\ll T_{\nu}(z=100)$, roughly  $m_{\phi}\lesssim 10^{-3}$ eV.
    They found the bound $g_{\nu,{\rm eff}}<2\times10^{-7}$, where the ratio between $g_{\nu, {\rm eff}}$ and $g_{\nu}$ is no larger than one order of magnitude.
    Finally, in the scalar mass region $10^{-1} {\rm eV}\lesssim m_{\phi}\lesssim 10^{3}{\rm eV}$ and using CMB data, a fairly robust neutrino-scalar bound was found by Escudero \& Witte \cite{Escudero2020_CMB_search}.
    In Fig. \ref{fig:mphi_vs_gnu}, we depicted all these bounds including our new measurement.

    We now discuss the bound on the electron-scalar coupling $g_e$.
    Notice that here we cannot set a direct constraint to $g_e$ because we do not have a direct measurement on $g_{\nu}$, we only have an upper bound. 
    We can only estimate where the bound would lie by using 
    the constraint on $G_{{\rm S}}$ together with the bound on $G_{{\rm eff}}$ from Eq.~\eqref{Gbounds}.
    Taking the $g_{\nu}$ upper value given in Eq.~\eqref{eq:gnu} we estimate
    \begin{equation}\label{eq:ge}
        g_e<2.7 \times 10^{-10} \left(\frac{m_{\phi}}{{\rm eV}} \right)~.
    \end{equation}
    
\begin{figure}
    \centering
    \includegraphics[width=0.47
    \textwidth]{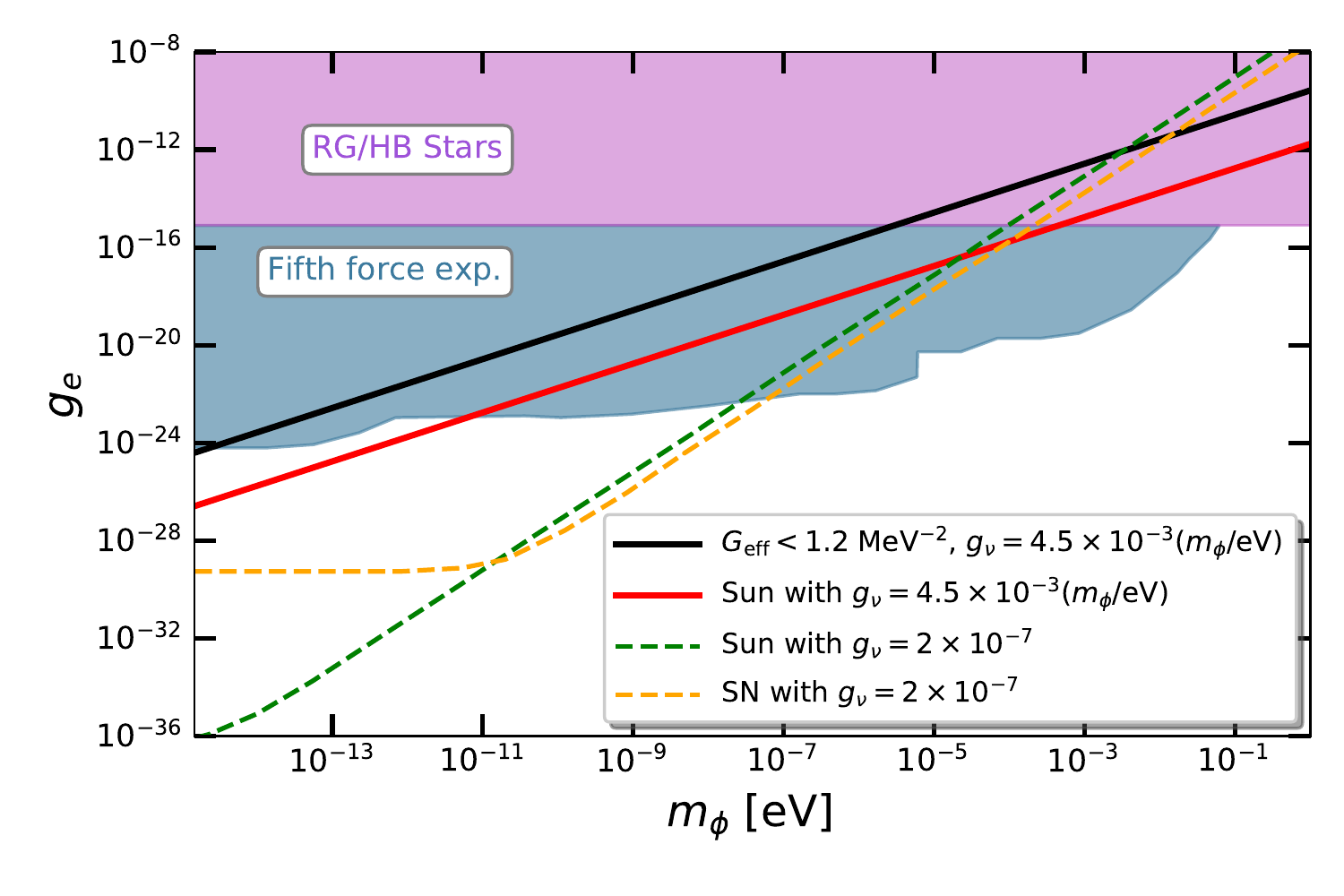}
    \caption{\label{fig:mphi_vs_ge}Electron-scalar coupling constraints.
    The region that lies above the black line is not permitted by our constraints for $g_{\nu}$-value fixed to its upper bound.
    The region above the red line would be prohibited for $g_{\nu}$-value fixed to its upper bound and the solar neutrino constraint.
    The purple area represents the region excluded by energy loss in stars.
    The blue area represents the region prohibited by fifth-force experiments.
    Dash lines: SN and Sun bounds by Babu \textit{et al.} 2020 for $g_{\nu}=2\times 10^{-7}$.}
    \end{figure}

    Supposing an eventual future measurement of $g_{\nu}$, we observe that the bound on $g_e$ is weaker in our analysis than those obtained from neutrinos SNSI from The Sun and SN.
    The Sun bound is particularly interesting, since \cite{Ge2019} found a possible preference for a non-vanishing SNSI $G_{{\rm eff}\odot}$ effective coupling.
    As a matter of fact, aside from the fluctuation, we can safely take the solar bound as $\Delta m_{\odot}<7.4\times 10^{-3}$ eV.
    The solar medium is non-relativistic, therefore, the mass correction in The Sun goes as  $\Delta m_{\odot}=G_{{\rm eff}\odot}n_{{\rm e \odot}}$, where the number density of electrons at the solar core is $n_{{\rm e \odot}}\sim 5.2 \times 10^{11}$ eV$^3$.
    With this $g_e=m_{\phi}^2\Delta m_{\odot}/g_{\nu}$.
    Comparing the solar bound on $\Delta m_{\odot}$ with our Early Universe bound, and taking the upper value in eq. \eqref{eq:gnu}, we observe that, indeed, the Early Universe bound is weaker than the solar one (see Fig. \ref{fig:mphi_vs_ge}).
    
    In Fig. \ref{fig:mphi_vs_ge} we show different constraints compared with our results.
    On the one hand side, there are strong bounds, $g_e<10^{-15}$, from stellar physics where an electron-scalar coupling would diminish stars to a cooler than expected state.
    This is due to energy loss caused by the unopposed escape of scalar particles produced from the stellar nucleus \cite{Hardy2017,Knapen2017}.
    On the other hand, fifth force experiments, that search for deviations to the Newtonian gravity set the strongest bounds for electron-scalar coupling with a very light mediator.
    The length scale of the gravitational experiments is related to the force mediator mass, which constraints can be directly obtained from \cite{Adelberger2009}.
    Experimental constraints at shorter lengths were reviewed and summarized by \cite{Babu2020scalar} with the results of several experiments \cite{Hoskins1985,*Long2003,*Kapner2007,*Geraci2008,*Yang2012,*Tan2016}.
    Here, we depict the compendium of fifth force experiments within a single bound in Fig. \ref{fig:mphi_vs_ge}.
    Additionally, we include the curves indicating the reference bound on $g_e$ for a fixed $g_{\nu}$-value in the case of neutrinos propagating in supernovas and The Sun from \cite{Babu2020scalar}.

    \section{\label{sec:conclusions}Summary and conclusions}

    We have performed a robust analysis of the consequences of a possible large neutrino effective mass due to thermal corrections mediated by nonstandard light scalar interactions among leptons in the context of the Early Universe.
    Such an effective mass is fed by CFS of propagating neutrinos through a thermal bath of neutrinos and electrons/positrons within the primordial plasma.
    At one-loop order, the effective neutrino mass is simply proportional to the respective scalar to neutrino/electron couplings, but inversely proportional to the square scalar mass. One can encode such dependencies in a couple of  SNSI parameters, $G_{\rm S}$ and $G_{\rm eff}$. The effective neutrino mass also depends on the temperature of the corresponding thermal bath, through a monotonically increasing function, such that, as higher the temperatures the larger the effective mass contributions. Hence, even if none visible effects appear at small redshifts,  possible changes on standard physics could arise as we look towards earlier times.
    
    In the case where the neutrino effective mass gets comparable with neutrino temperature, their number and energy density drops significantly. However, the SNSI effect vanishes faster than the temperature drop and, in equilibrium, the standard neutrino density is recovered.
    Nevertheless, once neutrinos decouple from the primordial plasma its production gets largely suppressed, thus, their density at decoupling freezes out. 
    This has an observable direct effect that is expressed as a smaller $N_{{\rm eff}}$ than expected. 
    
    BBN has shown to be sensitive to any nonstandard physics that affects the expansion rate. We have exploited this feature and used BBN primordial nuclei outputs and observational data to set a constraint on the neutrino-scalar coupling.
    Our new bound on $g_{\nu}$ is more restrictive that previously known bounds for the mass range $1.5 \times 10^{-15} {\rm eV}\lesssim m_{\phi}\lesssim 4.5 \times 10^{-5}{\rm eV}$.
    
    Although our analysis is able to constraint the scalar-electron couplings, it does also involve scalar-neutrino coupling, and, thus, no straightforward bound to the former can be set without knowledge about the latter.  Nevertheless, we have explored the parameter space assuming the saturation of our bound on the scalar-neutrino coupling to compare with other results from astrophysics and fifth force experimental limits.
    
    Along with our analysis, we have assumed that the light scalar mediator would play no direct role in early cosmology, by looking upon the parameter range where it would stay out of equilibrium, and its production rate suppressed during the Early Universe. 
    In the opposite scenario, neutrino NSI may have other consequences that can be further studied in cosmology.
    For instance, neutrino NSI may trigger active neutrino decays and annihilation into light bosons.
    Adding such effects to our analysis would probably amount to soften our bounds, since light scalars add to the relativistic degrees of freedom, rising $N_{\rm eff}$ and compensating the effect of thermal neutrino mass \cite{Huang2018,Luo2020}. 
    Furthermore, neutrino decay and annihilation during structure formation era could relax the bound on $\Sigma m_{\nu}$ from LSS \cite{Beacom2004,Hannestad2005,Escudero2019,Chacko2020,Escudero2020relaxing}, in contrast, larger bare neutrino masses impose more stringent constraints.
    Such analysis may be worthy of being pursued.
    
    The physics of nonstandard neutrino interactions is an active field of study due to its potential to solve current tensions in cosmology. The Early Universe can be used as a testing ground to study such interactions in environments unreachable by terrestrial or solar experiments. In this work, we used the indirect effect of neutrinos on the relic densities of light elements to impose bounds upon the possible interactions with a light scalar mediator. This bound is stronger than the previous bounds and contributes to a better understanding of the nature of neutrinos and its possible links to physics outside the standard model of particles.

    \begin{acknowledgments}
    This work has been partially supported by Conacyt, Mexico, under FORDECYT-PRONACES grant No. 490769.
    We thank two anonymous referees for their critical reviews that led to an improvement of our paper.
    \end{acknowledgments}
    
    \bibliography{references}
    
    \end{document}